\documentclass[sigconf]{acmart}
\usepackage{stfloats}

\setcopyright{rightsretained}

\AtBeginDocument{%
  \providecommand\BibTeX{{%
    \normalfont B\kern-0.5em{\scshape i\kern-0.25em b}\kern-0.8em\TeX}}}

\copyrightyear{2021} 
\acmYear{2021} 
\setcopyright{rightsretained} 
\acmConference[CHI '21 Extended Abstracts]{CHI Conference on Human Factors in Computing Systems Extended Abstracts}{May 8--13, 2021}{Yokohama, Japan}
\acmBooktitle{CHI Conference on Human Factors in Computing Systems Extended Abstracts (CHI '21 Extended Abstracts), May 8--13, 2021, Yokohama, Japan}\acmDOI{10.1145/3411763.3451985}
\acmISBN{978-1-4503-8095-9/21/05}

\begin{document}

\title[vrCAPTCHA]{vrCAPTCHA: Exploring CAPTCHA Designs in Virtual Reality}


\author{Xiang Li$^{1,2}$}
\affiliation{%
  \institution{$^1$Exertion Games Lab, Department of Human-Centred Computing, Monash University}
  \city{Melbourne}
  \country{Australia\\}
  \institution{$^2$Xi'an Jiaotong-Liverpool University}
  \city{Suzhou}
  \country{China}
}
\email{xiang.li18@student.xjtlu.edu.cn}

\author{Yuzheng Chen$^{1,2}$}
\affiliation{%
  \institution{$^1$Exertion Games Lab, Department of Human-Centred Computing, Monash University}
  \city{Melbourne}
  \country{Australia\\}
  \institution{$^2$Xi'an Jiaotong-Liverpool University}
  \city{Suzhou}
  \country{China}
}
\email{yuzheng.chen18@student.xjtlu.edu.cn}

\author{Rakesh Patibanda$^1$}
\affiliation{%
  \institution{$^1$Exertion Games Lab, Department of Human-Centred Computing, Monash University}
  \city{Melbourne}
  \country{Australia}
}
\email{rakesh@exertiongameslab.org}

\author{Florian 'Floyd' Mueller$^1$}
\authornote{Corresponding author.}
\affiliation{%
  \institution{$^1$Exertion Games Lab, Department of Human-Centred Computing, Monash University}
  \city{Melbourne}
  \country{Australia}
}
\email{floyd@exertiongameslab.org}

\renewcommand{\shortauthors}{Li, et al.}

\begin{abstract}
With the popularity of online access in virtual reality (VR) devices, it will become important to investigate exclusive and interactive CAPTCHA (Completely Automated Public Turing test to tell Computers and Humans Apart) designs for VR devices. In this paper, we first present four traditional two-dimensional (2D) CAPTCHAs (i.e., text-based, image-rotated, image-puzzled, and image-selected CAPTCHAs) in VR. Then, based on the three-dimensional (3D) interaction characteristics of VR devices, we propose two vrCAPTCHA design prototypes (i.e., task-driven and bodily motion-based CAPTCHAs). We conducted a user study with six participants for exploring the feasibility of our two vrCAPTCHAs and traditional CAPTCHAs in VR. We believe that our two vrCAPTCHAs can be an inspiration for the further design of CAPTCHAs in VR.
\end{abstract}




\maketitle

\section{Introduction}
CAPTCHAs (Completely Automated Public Turing test to tell Computers and Humans Apart) are used to protect the security of data \cite{VonAhn2008}. These authentication protocols can allow users to access websites while preventing automatic bots. The most common type of CAPTCHA is a series of graphical transformations on a two-dimensional (2D) screen, which makes it is nearly unsolvable by to an automatic bot. Recently, most mobile phones and computers support various CAPTCHAs to ensure user's access to websites. However, with the development of three-dimensional (3D) computing technology such as Virtual Reality (VR) headsets, user's pursuit of 2D space has moved to 3D environments, where we use mid-air \cite{koutsabasis2019empirical} gestures \cite{Jospe2017} to interact in 3D space and explore more immersive experiences.

In this paper, we present six CAPTCHA designs in VR, including four traditional CAPTCHAs (i.e., text-based, image-rotated, image-puzzled, and image-selected CAPTCHAs) and two interactive vrCAPTCHAs (i.e., task-driven and bodily motion-based CAPTCHAs). The aim is to explore the future CAPTCHA designs in a VR environment through leveraging the user's full physical interactions. We argue that the use of CAPTCHAs with entertaining interactivity features will allow users to immerse themselves in 3D. Besides, we believe that research on new virtual reality CAPTCHAs will become more important due to the proliferation of virtual reality devices, and we propose that our work can serve as a useful springboard for such future investigations. 

\section{Related Work}
\subsection{Traditional CAPTCHAs}

Constrained by the structure of 2D displays, traditional CAPTCHAs are almost always based on a 2D planar design. According to Feng et al. \cite{10.1145/3397312}, traditional 2D CAPTCHAs can be divided into two types: (1) text-based and (2) image-based CAPTCHAs.

\subsubsection{Text-based CAPTCHAs}

It is generally believed that the first practical text-based CAPTCHA was invented by Lillibridge \cite{Lillibridge2001} in 1997. Then, von Ahn et al. \cite{VonAhn2004} proposed reCAPTCHA, an authentication protocol that requires users to transcribe scanned text. As this is increasingly easier for bots, more recent research has tried to make OCR more difficult so that the bots cannot recognize the texts. For example, Baird et al. \cite{Baird2005} proposed ScatterType CAPTCHA, which prevents the computer from splitting characters from a word. However, the feasibility of text-input CAPTCHA in VR environments has not been fully discussed. Based on previous research on inputting text in VR \cite{zhai2005search}, it may be a misconception to still opt for text-based CAPTCHAs in VR.

\subsubsection{Image-based CAPTCHAs}
Bongo \cite{Golubitsky1985} was one of the first visual pattern recognition problems for CAPTCHAs. Chew et al. \cite{Chew2004} were among the first to work on tagging image-based CAPTCHAs. After that, many algorithms based on image recognition were proposed and developed. Asirra \cite{Elson2007} challenged the user in selecting images with cats among twelve images of cats and dogs. Matthews et al. \cite{Matthews2010} presented the scene marker test, where users must recognize relationships between irregularly shaped objects embedded in a background image. However, the flat image-based CAPTCHA mainly considers devices with a 2D display plane as the carrier, and the interactive nature of 3D virtual objects unique to VR has not been fully investigated.

\subsection{Mid-air Interaction}
Koutsabasis and Vogiatzidakis \cite{koutsabasis2019empirical} pointed out that mid-air interactions should have the following three characteristics: (1) touchless interaction, (2) real-time sensors that can track some of the user's physical activity and movement, and (3) body movements, postures, and gestures need to be recognized and matched to specific user intents, goals and commands \cite{Xu2020b}. However, traditional CAPTCHAs are difficult to create such mid-air interactions environment in VR. Since most traditional CAPTCHAs rely on horizontal and vertical planes of interaction using mouse clicks and touch screen 
taps. We conjecture that these types of interaction are not applicable to the user experience of interacting with CAPTCHAs in VR. Therefore, we selected four traditional text- and image-based CAPTCHAs and implemented them in a virtual reality environment as prototypes for a preliminary study to examine the associated user experience.

\section{DESIGN AND DEMONSTRATION}

\begin{figure*}[htp]
  \centering
  \includegraphics[width=\linewidth]{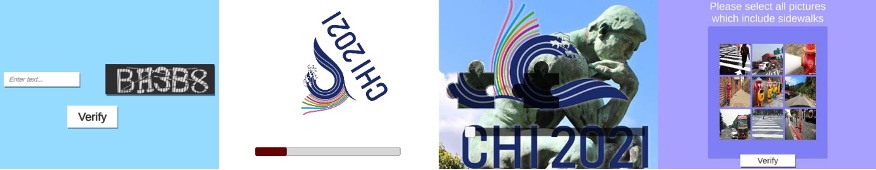}
  \caption{Four traditional CAPTCHAs: (1) Text-based CAPTCHA; (2) Image-rotated CAPTCHA; (3) Image-puzzled CAPTCHA; and (4) Image-selected CAPTCHA.}
  \Description{Selected four traditional CAPTCHAs.}
\end{figure*}

\subsection{Four Traditional CAPTCHAs in VR}
To guide our design, we firstly considered four traditional CAPTCHAs (i.e., text-based, image-rotated, image-puzzled, and image-selected CAPTCHAs).

\subsubsection{Text-based CAPTCHA}
In Figure 1.1, the participants were asked to use the trigger button of the right controller to click on the blank input field for text input to bring up the virtual keyboard. Participants then used the VR controller to click the corresponding letters or numbers and finally clicked the verify button.

\subsubsection{Image-rotated CAPTCHA}
In Figure 1.2, the participants were provided with a tilted image. Then, the participant was asked to drag the slider on the progress bar by holding the trigger button on the controller. The movement of this slider would change the rotation of the image. Once the participant thought that the rotation angle of the image had reached the correct position, they should release the trigger button to verify it.

\subsubsection{Image-puzzled CAPTCHA}
In Figure 1.3, the participants were asked to drag the slider by holding the trigger button on the right controller. Unlike the previous technique, the movement of the handle changed the horizontal position of the puzzle piece. Once the participant believed that the puzzle piece had been stitched to the original image, they should release the trigger button.

\subsubsection{Image-selected CAPTCHA}
In Figure 1.4, the participants were asked to follow the text to select all of the nine images that meet the requirements. Participants selected images by moving the slider along the controllers and pressing the trigger button when they finished their selections. When all images were selected, they were asked to press the verify button.

\subsection{Early Exploration of vrCAPTCHAs}
\subsubsection{Task-driven CAPTCHA}

\begin{figure}[h]
  \centering
  \includegraphics[width=\linewidth]{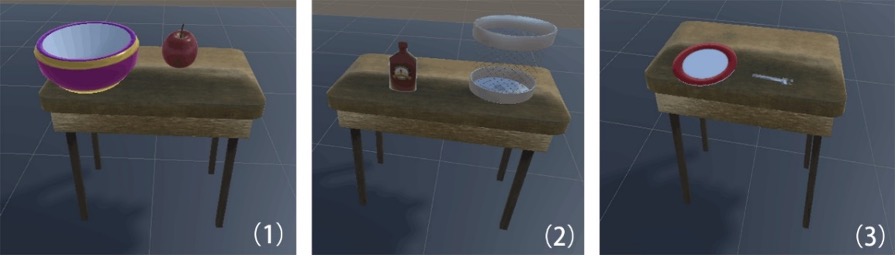}
  \caption{Take-driven CAPTCHA.}
  \Description{Take-driven CAPTCHA.}
\end{figure}

We first designed a CAPTCHA that requires the user to perform a simple action to authenticate. This CAPTCHA requires the user to lift a virtual object (e.g., apple, wine bottle or silver knife and fork) by holding the trigger button and moving it to a specified location (e.g., purple bowl (Fig. 2.1), trash can (Fig. 2.2), or dinner plate (Fig. 2.3)). This task-driven CAPTCHA is based on the participant’s feedback. “I think you can design a real-life scenario to enhance the interaction of CAPTCHA.” (P5).

\subsubsection{Motion-based CAPTCHA}

\begin{figure}[t]
  \centering
  \includegraphics[scale=0.7]{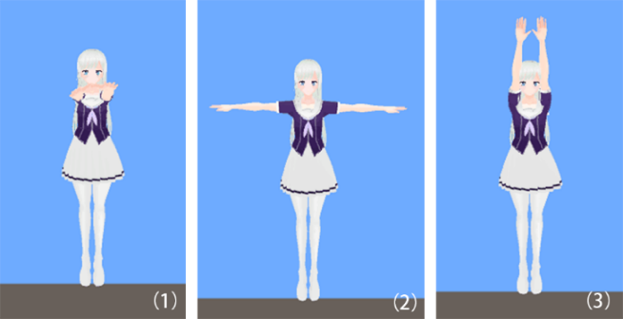}
  \caption{Motion-based CAPTCHA.}
  \Description{Motion-based CAPTCHA.}
\end{figure}

We also considered using bodily motions to implement authentication interaction, and we named it motion-based CAPTCHA. With this CAPTCHA, the users are asked to follow the movements of a virtual character that appears in front of them (e.g., body movement like front flat raise (Fig. 3.1), side flat raise (Fig. 3.2) and up raise (Fig. 3.3)). This motion-based CAPTCHA is inspired by previous exergame research studies \cite{Li2020, xu2020virusboxing, Xu2020, Arnold2018, Patibanda2017} which demonstrate the immersion that upper body limb movements can bring in VR. It is engaging for users to follow the virtual character and move their bodies.

\section{Pre-Study and Feedback}
We recruited six participants to finish the pre-user study for these four traditional CAPTCHAs and our two new vrCAPTCHAs. According to our subjective ranking feedback from participants, all participants ranked the task-driven CAPTCHA as their favourite. “I think it is the most interesting CAPTCHA!” (P1). “It is very interesting; I enjoy using this CAPTCHA in VR.” (P6). Four participants thought that the motion-based CAPTCHA is the second best. “I like this (motion-based) CAPTCHA, it makes me feel that I am playing a VR game.” (P2, P4). All participants thought that the two vrCAPTCHAs could greatly improve their motivation to experience CAPTCHA in VR. Four participants ranked the text-based CAPTCHA last in their preferences. For example, P1 complained that the “text-based CAPTCHA was too time consuming and cumbersome in VR.”

\section{Limitations and Discussion}
We acknowledge the limitations of our two vrCAPTCHAs. For task-driven CAPTCHAs, we only considered simple movements of objects. In fact, for virtual 3D spaces, we can design richer interactive environments, such as simulating daily activities like playing basketball or pouring tea. For motion-based CAPTCHAs, we considered only the top half of the user’s body. This can be done by introducing other sensors to enable full body interaction and thus provide fuller immersion, for example, using a Kinect camera. 

We also acknowledge that the security of our vrCAPTCHAs is insufficiently validated. Given that, we here provide a discussion about interactive vrCAPTCHAs. Unlike the traditional CAPTCHAs, previous CAPTCHAs are verified by computational processing of images or text that cannot be recognized by computers. Our CAPTCHA requires the user to follow the CAPTCHA to perform the relevant physical actions to complete the submission of the CAPTCHA. Besides, according to Mueller et al. \cite{Mueller2001}, our subsequent work in the future will focus on proposing fuzziness validation tests based on the characteristics exhibited by the user as a human interaction, thus distinguishing the concept of absolute accuracy achieved by the computer.

\section{CONCLUSION}
In this paper, we propose two interactive vrCAPTCHAs (i.e., task-driven and bodily motion-driven CAPTCHAs), inspired by the previous 2D traditional CAPTCHA designs (i.e., text-based, image-rotated, image-puzzled, and image-selected CAPTCHAs) and combined with the 3D interactivity of VR. We conducted a user study with six participants to evaluate the feasibility of traditional CAPTC-HAs and vrCAPTCHAs in VR. We believe that our two vrCAPTCHAs can be an inspiration for the future design of CAPTCHAs in VR.

\begin{acks}
We are grateful to the six volunteers who were willing to come and participate in our pre-user study. Xiang Li would like to thank his internship advisors Prof. Florian ‘Floyd’ Mueller and Rakesh Patibanda, and other members at the Exertion Games Lab of Monash University for their valuable feedback during the prototyping of this work. 
\end{acks}

\bibliographystyle{ACM-Reference-Format}
\bibliography{vrCAPTCHA}


\end{document}